\documentclass[]{aa}

\usepackage{graphicx}
\usepackage{txfonts}
\usepackage{url}
\usepackage{natbib}
\usepackage{color}
\usepackage{textcomp}
\usepackage{gensymb}
\bibpunct{(}{)}{;}{a}{}{,}
\usepackage{amstext}

\begin{document}

\title{Spinning nano-carbon grains: Viable origin for anomalous microwave emission}

\author{N. Ysard\inst{\ref{inst1}}
\and M.-A. Miville-Desch{\^e}nes\inst{\ref{inst2}}
\and L. Verstraete\inst{\ref{inst1}}
\and A.P. Jones\inst{\ref{inst1}}
}

\institute{Universit{\'e} Paris-Saclay, CNRS, Institut d'astrophysique spatiale, 91405, Orsay, France\label{inst1}
\and AIM, CEA, CNRS, Universit{\'e} Paris-Saclay, Universit{\'e} Paris Diderot, Sorbonne Paris Cit{\'e}, 91191 Gif-sur-Yvette, France\label{inst2}\\
 \email{nathalie.ysard@universite-paris-saclay.fr}}
 
\abstract
{Excess microwave emission, commonly known as anomalous microwave emission (AME), is now routinely detected in the Milky Way. Although its link with the rotation of interstellar (carbonaceous) nano-grains seems to be relatively well established at cloud scales, large-scale observations show a lack of correlation between the different tracers of nano-carbons and AME, which has led the community to question the viability of this link.}
{Using ancillary data and spinning dust models for nano-carbons and nano-silicates, we explore the extent to which the AME that come out of the Galactic Plane might originate with one or another carrier.}
{In contrast to previous large-scale studies, our method is not built on comparing the correlations of the different dust tracers with each other, but rather on comparing the poor correlations predicted by the models with observed correlations. This is based on estimates that are as realistic as
possible of the gas ionisation state and grain charge as a function of the local radiation field and gas density.}
{First, nano-carbon dust can explain all the observations for medium properties, in agreement with the latest findings about the separation of cold and warm neutral medium in the diffuse interstellar medium. The dispersion in the observations can be accounted for with little variations in the dust size distribution, abundance, or electric dipole moment. Second, regardless of the properties and abundance of the nano-silicate dust we considered, spinning nano-silicates are excluded as the sole source of the AME. Third, the best agreement with the observations is obtained when the emission of spinning nano-carbons alone is taken into account. However, a marginal participation of nano-silicates in AME production cannot be excluded as long as their abundance does not exceed $Y_{\rm Si} \sim 1\%$.}
{}

   \authorrunning{}
\titlerunning{}
\maketitle

\section{Introduction}
\label{introduction}

First detected in the 1990s as a dust-correlated excess component above the free-free and synchroton emissions \citep[e.g.][]{Kogut1996, Leitch1997}, anomalous microwave emission (AME) has since been observed to be bright in the Milky Way and other galaxies \citep[see the review by][]{Dickinson2018}, with coherent AME structures from scales of degrees down to arcseconds. The prevalent explanation is that the AME carriers are spinning carbonaceous nano-grains emitting through spontaneous emission of purely rotational photons \citep[e.g.][]{DL98, Ali2009, Ysard2010, Ysard2010b, Silsbee2011}. Other carriers of nanometric size, such as spinning nano-silicates, may also contribute to AME, however \citep{Hoang2016, Hensley2017}. Other AME origins were also suggested, such as spinning-grain magnetic dipole emission \citep{DH2013, HL2016}, thermal emission from magnetic dust \citep{DL99} and variations in the sub-millimetre opacity of amorphous solids at low temperature \citep{Jones2009, Nashimoto2020}. All of these cannot entirely reconcile the low polarisation degree of the 30~GHz AME, which is lower than 1\% in the diffuse interstellar medium \citep[see Table~4 in][and references therein]{Dickinson2018}.

Anomalous microwave emission is a bright component that accounts for about half of the observed intensity at 30~GHz in the Galactic Plane \citep{Planck2014, PlanckCOMMANDERb, COMMANDER2016}. It is brightest in interstellar regions hosting photon-dominated regions \citep[PDRs;][among others]{Casassus2008, Casassus2021, Cepeda2021, Tibbs2010}. These highly irradiated transition regions between HII regions and molecular clouds are known to be sites of strong and rapid evolution of the sub-nanometre hydrocarbon grain populations. Many studies have shown that the abundance of the carbonaceous nano-grains decreases but their minimum size increases from the outer PDRs to the inner molecular clouds. This holds true regardless of the assumed type of carbonaceous nano-grains, that is, PAHs \citep[e.g.][]{Berne2007, Compiegne2008, Arab2012, Pilleri2012, Pilleri2015} or amorphous hydrocarbons a-C(:H) \citep{Schirmer2020}, and is taken to be the result of photoprocessing in all cases \citep[see e.g.][]{Murga2016, Murga2019}. \citet{Casassus2021} showed that this scenario is consistent with AME ATCA observations of the $\rho$ Oph W PDR at an angular resolution of 30''. They found that the peak frequency of the AME increases from the inner to the outer PDR and that the correlation with mid-IR emission in the Spitzer IRAC 8~$\mu$m filter is tight in the inner PDR, but less so towards the exciting star. Fitting these data with spinning PAHs was possible by adopting an increase in the minimum size deeper into the PDR, in agreement with previous PDR studies based on near- to far-IR observations. As stated by \citet{Casassus2021}, at small scales, the AME appears to correlate tightly with the carbonaceous nano-grain mid-IR emission. This was also shown to be the case for other PDRs \citep[e.g.][]{Casassus2008, Tibbs2010, Tibbs2011, Scaife2010, Bell2019, Cepeda2021} and the dark clouds LDN1622 and LDN1780 \citep[][respectively]{Harper2015, Vidal2020}. \citet{Vidal2020} found that the observed variations in the 30~GHz emissivity observed in LDN1780 are consistent with spinning PAHs, with size distribution variations from the outer layers to the dense core of the molecular cloud. This agrees with the 3D radiative transfer modelling of IR data performed by \citet{Ridderstad2006}.

These small-scale studies clearly illustrate that despite the good agreement between the spinning carbonaceous nano-grain emission model and the observed AME, the correlation between the mid-IR emission and the AME can be rather poor. This is especially true in lines of sight that are dominated by high-density or high-radiation field bright regions, for example, throughout the Galactic Plane. The starting point of our study is to address the challenge to the validity of the rotating carbonaceous nano-grain origin of the AME by \citet{Hensley2016}. They performed a full-sky analysis at a 1$\degree$ angular resolution. Their conclusions were based on the poor correlation of the 30~GHz AME with the observed mid-IR emission and on the way the AME correlates with the dust far-IR to submm radiance. They concluded that other carriers, such as nano-silicates, or other emission mechanisms had to be considered. Using the same AME data as \citet{Hensley2016}, we perform a full-sky analysis with the aim of reconciling the small- and large-scale observations. However, our method does not compare the correlations between the different dust tracers, but, in agreement with small-scale studies, compares the poor correlations predicted by the models with observed correlations. In order to be comprehensive, we consider both spinning nano-silicate and nano-carbon dust emission.

The paper is organised as follows. Section~\ref{observations} describes the data sets and Sect.~\ref{models} the dust properties and spinning-dust model. We then discuss the latest observational findings for the density of the neutral gas and its distribution in the diffuse interstellar medium in Sect.~\ref{CNM_vs_WNM}. Our results are presented in Sect.~\ref{models_vs_observations}, and we conclude in Sect.~\ref{conclusion}.

\section{Observational data}
\label{observations}

In addition to maps of the AME, we used data that are characteristic of the thermal emission of grains, both large and small, to carry out our analysis. We also used data characteristic of the radiation field and the column density.

\subsection{Planck LFI and HFI foreground products}
\label{planck_lfi}

The Planck observatory\footnote{Planck (\url{http://www.esa.int/Planck}) is a project of the European Space Agency (ESA) with instruments provided by two scientific consortia funded by ESA member states and led by Principal Investigators from France and Italy, telescope reflectors provided through a collaboration between ESA and a scientific consortium led and funded by Denmark, and additional contributions from NASA (USA).} carried out a full-sky survey in nine frequency channels between 30 and 857~GHz, that is, $\sim 1$~cm and 350~$\mu$m. This was subsequently used to estimate the diffuse Galactic emission components: AME and dust thermal emission, and free-free and synchrotron emissions.

For the AME, we used the decomposition made by \citet{COMMANDER2016}, which combined the Planck temperature maps in all channels of the Low Frequency Instrument (LFI: 30, 44 and 70~GHz) and High Frequency Instrument (HFI: 100, 143, 353 and 857~GHz), the 9-year sky maps of the Wilkinson Microwave Anisotropy Probe \citep[WMAP: five channels from 23 to 94~GHz;][]{Bennett2013}, and the \citet{Haslam1982} 408~MHz map. This allowed \citet{COMMANDER2016} to deliver a parametric model of AME at an angular resolution of 1$\degree$. The model parameters are available as a healpix map \citep{Gorski2005} with $n_{\rm side} = 256$ that can be downloaded from the Planck Legacy Archive\footnote{\url{http://pla.esac.esa.int/pla/}.} (PLA). With these, we computed AME maps at 20, 30, and 40~GHz. 

\citet{PlanckCollaborationXI} combined the Planck HFI maps with the IRIS 100~$\mu$m map \citep{MAMD2005}. Performing a pixel-by-pixel modified blackbody $\chi^2$ fit, \citet{PlanckCollaborationXI} derived the dust far-IR radiance, $R = \int_\nu I_{\nu} d\nu$, and optical depth at 353~GHz, $\tau_{353{\rm GHz}}$. These product maps can be downloaded from the PLA as healpix maps with a 1$\degree$ angular resolution and $n_{\rm side} = 256$. The radiance is useful as a good proxy for the radiation field, and the optical depth gives an estimate of the emissivity of large grains that in principle is independent of temperature. Both of these assertions are questionable in the dense molecular regions, however, were large variations in the temperature along the line of sight and variations in the dust grain optical properties are expected \citep[e.g.][]{PlanckCollaborationXI}.

\subsection{Thermal dust emission}
\label{thermal_dust_emission}

To characterise the large-grain thermal emission, we used the 100~$\mu$m map described in \citet{PlanckCollaborationXI}. It is a combination of the map of \citet{Schlegel1998} at scales larger than 30' and of the IRIS map of \citet{MAMD2005} at smaller scales. This combination provides the large-scale structure of the maps of \citet{Schlegel1998}, which have a higher quality because they are less contaminated by zodiacal light residuals, together with the higher resolution and better calibration of the IRIS map \citep[see Sect. 2.2 in][for details]{PlanckCollaborationXI}. The IRIS maps are derived from the IRAS data and have an angular resolution of 5' and central wavelengths of 12, 25, 60, and 100~$\mu$m with bandwidths of $\sim 7.0, 11.15, 32.5,$ and 31.5~$\mu$m \citep{IRAS, MAMD2005}.

In addition, the nano-grain thermal emission was characterised using the 12~$\mu$m IRIS map \citep{MAMD2005}, from which zodiacal residuals were filtered out using a matched filter technique tailored to extracting diffuse emission correlated to the ecliptic reference frame. The technique is described in Appendix~\ref{IRIS}. The 12 and 100~$\mu$m IRIS maps were smoothed to a full width at half maximum, FWHM = $1^\circ$, and projected on a $n_{\rm side} = 256$ healpix grid.

\subsection{Column density map}
\label{column_density}

As stated in Sect.~\ref{introduction}, our premise is that AME is produced by the rotation of (sub-)nanometre-sized grains that radiate most of their energy in the near- to mid-IR. It has been known for decades that the dust SED in this spectral range varies strongly depending on the observed interstellar regions \citep[e.g.][]{Laureijs1991, Abergel1996, Berne2007}. These variations are commonly interpreted in terms of variations in the size distribution and abundance of nano-grains \citep[see][for AME related examples]{Ridderstad2006, Vidal2020, Casassus2021}. In particular, there is a deficit in the mid-IR relative to the far-IR emission at atomic diffuse lines of sight compared to lines of sight containing molecular clouds. This can be explained by the disappearance of the small nano-grains together with the growth of the larger grains by grain-grain coagulation \citep[e.g.][]{Bernard1999, Stepnik2003, Flagey2009, Ysard2013}. This disappearance is in agreement with the variations observed in the 2175~$\mathring{A}$ bump towards molecular clouds \citep[e.g.][]{Mathis1989, Kim1994, Campeggio2007}, while the growth of the larger grains can explain the observed increase in selective extinction $R_V$ \citep[e.g.][]{Ormel2009, Koehler2015}.

Furthermore, regardless of their specifics in terms of composition and initial grain size distribution, all models indicate that the disappearance of the nano-grains occurs and the growth of larger grains starts to occur at still relatively low densities of the order of (a few) $10^3$~H/cm$^3$. This has been shown from observations of (i) the thermal emission of grains from the mid- to far-IR \citep[e.g.][]{Ridderstad2006, Ysard2013, Saajasto2021}, (ii) the near- to mid-IR scattering known as cloudshine and coreshine \citep[e.g.][]{Lefevre2014, Ysard2016, Juvela2020}, and (iii) the polarisation \citep[e.g.][]{Fanciullo2017, Vaillancourt2020}. Depending on the dust model details and the gas local properties, the accretion of the (sub-)nanometre-sized grains on larger grains is not expected to take more than a hundred to a few thousand years in the outskirts of molecular clouds \citep[e.g.][]{Stepnik2003, Koehler2012, Ysard2013, Jones2014}. Therefore, when we start from the assumption that nano-grains cause the AME, then the microwave flux must come primarily from the atomic diffuse interstellar medium, and the HI column density should be used when the AME and the 12~$\mu$m emission are modelled.

Consequently, the HI column density map we used is that of the HI4PI survey \citep{HI4PI2016}. It was constructed by adding all velocity channels between -90 and +90 km\,s$^{-1}$ to avoid emission from high-velocity clouds. The native angular resolution of this map is FWHM = 16.2'. It was then convolved with the required complementary Gaussian beam to bring the map to FWHM = $1^\circ$ before we reprojected to a $n_{\rm side} = 256$ healpix grid. 

\subsection{Masks}
\label{masks}

\begin{figure}[!t]
\centerline{\includegraphics[width=0.5\textwidth]{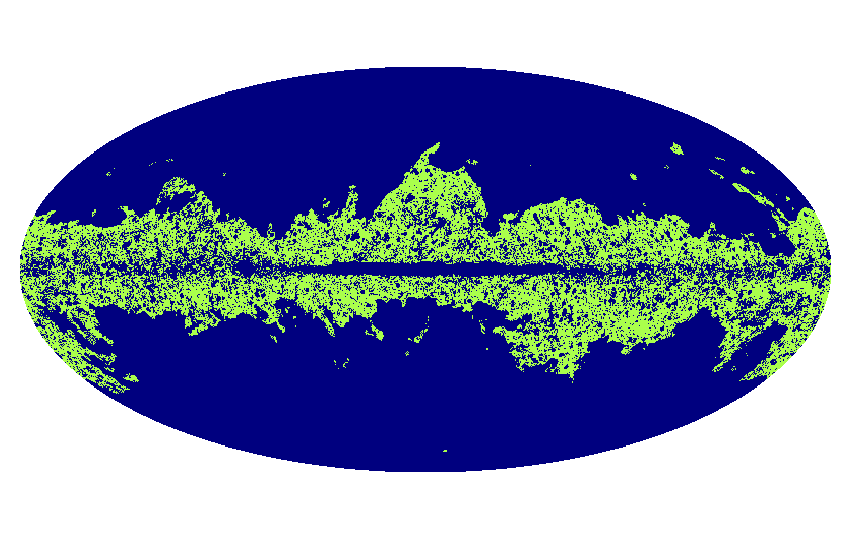}}
\caption{Mask used to make the analysis. We used the green pixels.}
\label{figure_mask} 
\end{figure}

\begin{figure}[!t]
\centerline{\includegraphics[width=0.5\textwidth]{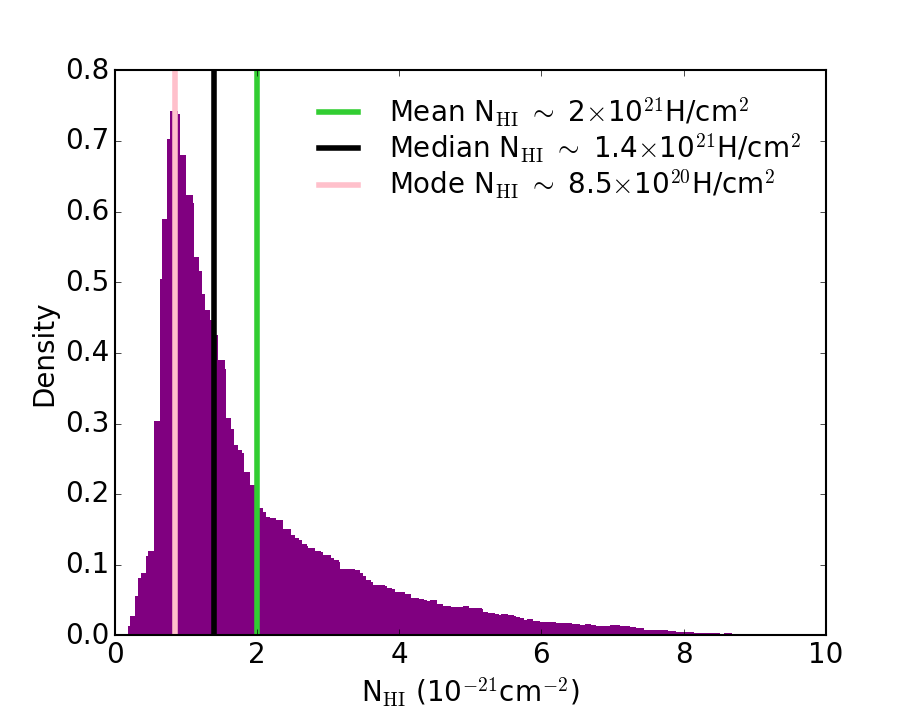}}
\caption{Histogram of the column density of the pixels we selected for the analysis.}
\label{figure_NHI_distribution} 
\end{figure}

In the following, we exclude a number of pixels from our analysis. First, as we focused on the neutral diffuse medium, we excluded the Galactic Plane, which contains many molecular regions as well as very bright HII complexes (e.g. Cygnus X) that are too complicated to be modelled globally. To do this, we used the mask defined by \citet{PlanckMasks}, based on the intensity at 353~GHz, excluding 1\% of the sky. We also excluded the point sources detected in the six Planck HFI frequency channels at 100, 143, 217, 353, 545, and 857~GHz and in the Planck LFI 30~GHz channel \citep{PlanckMasks}. This resulted in an exclusion of $\sim 23\%$ of the sky for healpix maps with $n_{side} = 256$. A third mask was applied to the data to exclude pixels for which the S/N~$< 3$ in the 30~GHz AME map. This led to the rejection of high Galactic latitude pixels for which the AME intensity was very low and covered less than 1\% of the sky. According to \citet{PlanckCOMMANDERb}, the component separation solution in these pixels is dominated by instrumental noise. Finally, we removed the sky area where the S/N~$< 3$ in the 12~$\mu$m IRIS map. The uncertainty in these mainly high-latitude areas is dominated by the uncertainty in the subtraction of the zodiacal emission. This amounts to eliminating $\sim 63\%$ of the sky. The resulting mask is displayed in Fig.~\ref{figure_mask}. It mainly excludes the Galactic Plane and the highest-latitude regions, that is, about 76\% of the sky.

This leads to data with column densities ranging from $\sim 10^{20}$ to $1.4 \times 10^{22}$~H/cm$^2$ (Fig.~\ref{figure_NHI_distribution}). The mean, median, and mode of the column density distribution are $\sim 2\times 10^{21}$, $1.4 \times 10^{21}$ , and $8.5 \times 10^{20}$~H/cm$^2$, respectively. The standard deviation is $\sim 1.5 \times 10^{21}$~H/cm$^2$.

\section{Models: Spinning and thermal dust emission}
\label{models}

All estimates presented in the following were obtained with the numerical code DustEM\footnote{Available here: \url{https://www.ias.u-psud.fr/DUSTEM/}.} \citep{Compiegne2011}. DustEM is a versatile numerical tool that is mainly used to calculate the extinction and thermal emission (polarised or unpolarised) of any grain model. A calculation of the spinning dust emission is also implemented. The specificity of this implementation is that it allows the user to process all the parameters involved in the calculation of the excitation and damping of the grain rotation in a consistent way according to their size and composition.

The rate of grain rotation depends on their emission of rovibrational and rotational photons, and on how they interact with the gas. To quantify these interactions, it is necessary to know the charge distribution of the grains and the main parameters describing the state of the gas. These parameters are its temperature $T_{gas}$, the electron abundance $x_e = n_e/n_{\rm H}$, the molecular fraction $y = n({\rm H}_2)/n_{\rm H}$ , and the ionisation fraction of the main species $x_{\rm H} = n({\rm HII})/n_{\rm H}, x_{\rm C} = n({\rm CII})/n_{\rm H}$ , where $n_{\rm H} = n({\rm HI}) + n({\rm HII})+2n({\rm H}_2$). The equilibrium charge distribution $f(Z)$ of grains per size and type was computed using the \citet{Kimura2016} model for the photoemission yield and the \citet{Weingartner2001} formalism for the remaining processes\footnote{Details about the model can be found here: \url{https://www.ias.u-psud.fr/DUSTEM/userguide.html}.}. In turn, the parameters describing the gas state were computed using the formalism described in \citet{Ysard2011}.

\subsection{Dust properties}
\label{dust_properties}

The dust properties we used are those of The Heterogeneous dust Evolution Model for Interstellar Solids \citep[THEMIS\footnote{Available here: \url{https://www.ias.u-psud.fr/themis/}.},][]{Jones2017}. It assumes that dust in the diffuse interstellar medium consists of small nano-grains of aromatic-rich carbon (radius $ 0.4 \leqslant a \leqslant 20$~nm) and larger core-mantle grains, for which the cores are composed either of amorphous aliphatic-rich carbon or of a mixture of amorphous silicates with the normative chemical compositions of olivine and pyroxene (with half of the total mass of silicate grains in each type). For all types of grains, the mantles are made of aromatic-rich carbons with a thickness of 20~nm for the carbon cores and 5~nm for the silicate cores. Moreover, iron and sulphur are incorporated into the silicate cores in the form of metallic nano-inclusions of Fe and FeS. They represent 10\% of the total core volume, of which 30\% is FeS and 70\% is pure Fe. This model allows reproducing the emission and extinction curves that are representative of the diffuse medium at high Galactic latitude from the optical to the sub-millimetre \citep[c.f.][]{Jones2013, Ysard2015}.

To model spinning nano-carbon dust emission, we used the default size distribution of THEMIS, which is a power law $\propto a^{-5}$ with an exponential cut-off and a minimum size of 0.4~nm. We assumed their permanent electric dipole moment to be $\mu \sim m \sqrt{N_{at}}$ , where $N_{at}$ is the number of atoms in the grain and $m = 0.38$~D \citep[see references in][]{Draine1998}. The THEMIS nano-carbon grains have \citep{Jones2012a, Jones2012b, Jones2012c}
\begin{equation}
N_{at} = 2500 \left(\frac{a}{1~\rm{nm}}\right) \left(\frac{\rho}{12 - 11X_{\rm H}}\right),
\end{equation}
where $X_{\rm H}$ is the grain hydrogen fraction and $\rho$ is its density, which are equal to 0.023 and 1.6~g/cm$^3$ for the nano-carbons of interest for rotational emission.

THEMIS does not include nano-silicates. To date, the observations have not provided any evidence of their presence in the diffuse medium. However, this lack of detection does not rule them out as a minority constituent, and other dust models include nano-silicates and have placed tentative upper limits on their abundance \citep[e.g.][]{Desert1986, Li2001}. In addition to the THEMIS dust components, we followed the studies of \citet{Hoang2016} and \citet{Hensley2017} and defined a size distribution for nano-silicates and set their electric dipole moment and abundance to model spinning
nano-silicate dust emission. To be able to explain AME by the rotation of nano-silicates without contradicting other observations, they cannot contain more than 14\% of the available silicon and their dipole moment per atom must be $0.2 \leqslant m \leqslant 1$~D, according to these two studies. To estimate the nano-silicate dipole moments, we assumed that they are a 50\%-50\% mixture of grains that have the normative compositions of forsterite and enstatite so that we were able to calculate the number of atoms as a function of their size. For the size distribution, both studies assumed log-normal distributions centred on a subnanometric size $a_0$ and of variable width $\sigma$. In contrast to \citet{Hoang2016}, \citet{Hensley2017} took into account the sublimation rates calculated by \citet{Guhathakurta1989} to set a minimum nano-silicate grain size. They found that when it is illuminated by the average interstellar radiation field, the critical survival size is of 37 atoms, that is, about 0.45~nm. For the models we present below, we define four cases, assuming that $Y_{\rm Si} = 10\%$ of the total silicon included in dust is locked in nano-silicates, the upper limit derived by \citet{Li2001}, and that $a_0 = a_{min} = 0.45$~nm as in \citet{Hensley2017}:
\begin{itemize}
\item case 1: $m = 0.3$~D and $\sigma = 0.3$ ;
\item case 2: $m = 0.3$~D and $\sigma = 0.1$ ;
\item case 3: $m = 1$~D and $\sigma = 0.3$ ;
\item case 4: $m = 1$~D and $\sigma = 0.1$.
\end{itemize}
This roughly covers the parameter space defined as acceptable by \citet{Hoang2016} and \citet{Hensley2017}. 

\subsection{Properties of the local medium }
\label{medium_properties}

To compare them with observations, model grids were generated for nano-carbons and nano-silicates. When the dust properties are settled, DustEM only requires three inputs: the radiation field, the gas density, and the gas temperature.

We assumed that the interstellar radiation field illuminating the grains has the same spectral distribution as the field of \citet{Mathis1983}. Its intensity was then scaled by the $G_0$ factor with $0.1 \leqslant G_0 \leqslant 20$ \text{and }$G_0 = 1$ corresponding to the \citet{Mathis1983} standard field intensity in the solar neighbourhood. For the gas, we started from a highly diffuse medium with $n_{\rm H} = 0.05$~H/cm$^3$ to reach densities that are representative of translucent clouds with $n_{\rm H} = 1\,000$~H/cm$^3$. This wide range of parameters should enable us a priori to account for the physical conditions of most lines of sight of the out of Galactic Plane that we considered. The corresponding gas temperatures were computed with CLOUDY \citep{Ferland1998}. The other gas parameters were directly calculated by DustEM \citep[see][for details]{Ysard2011}.

\begin{figure*}[!t]
\centerline{\begin{tabular}{cc}
\includegraphics[width=0.5\textwidth]{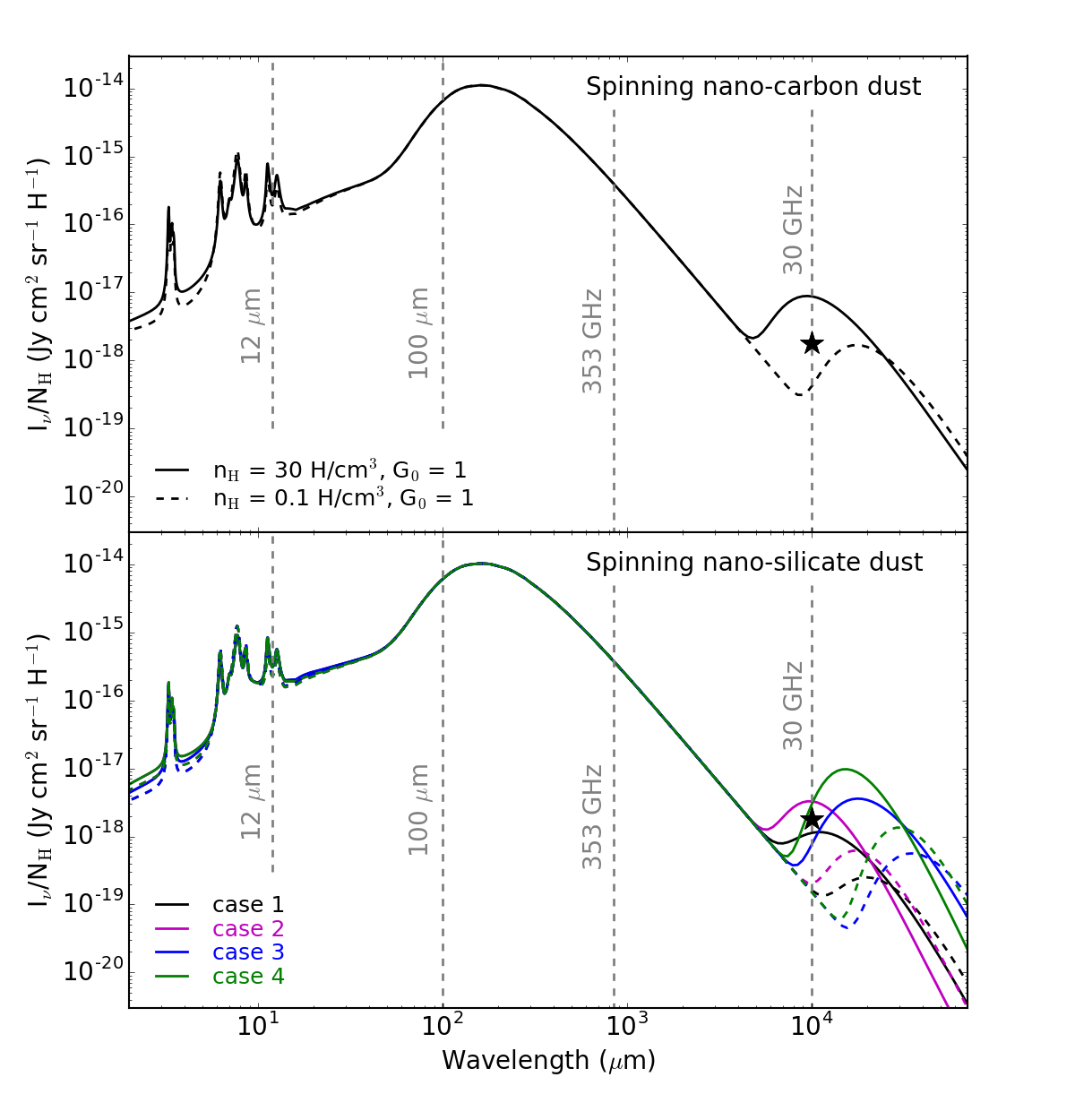} & \includegraphics[width=0.5\textwidth]{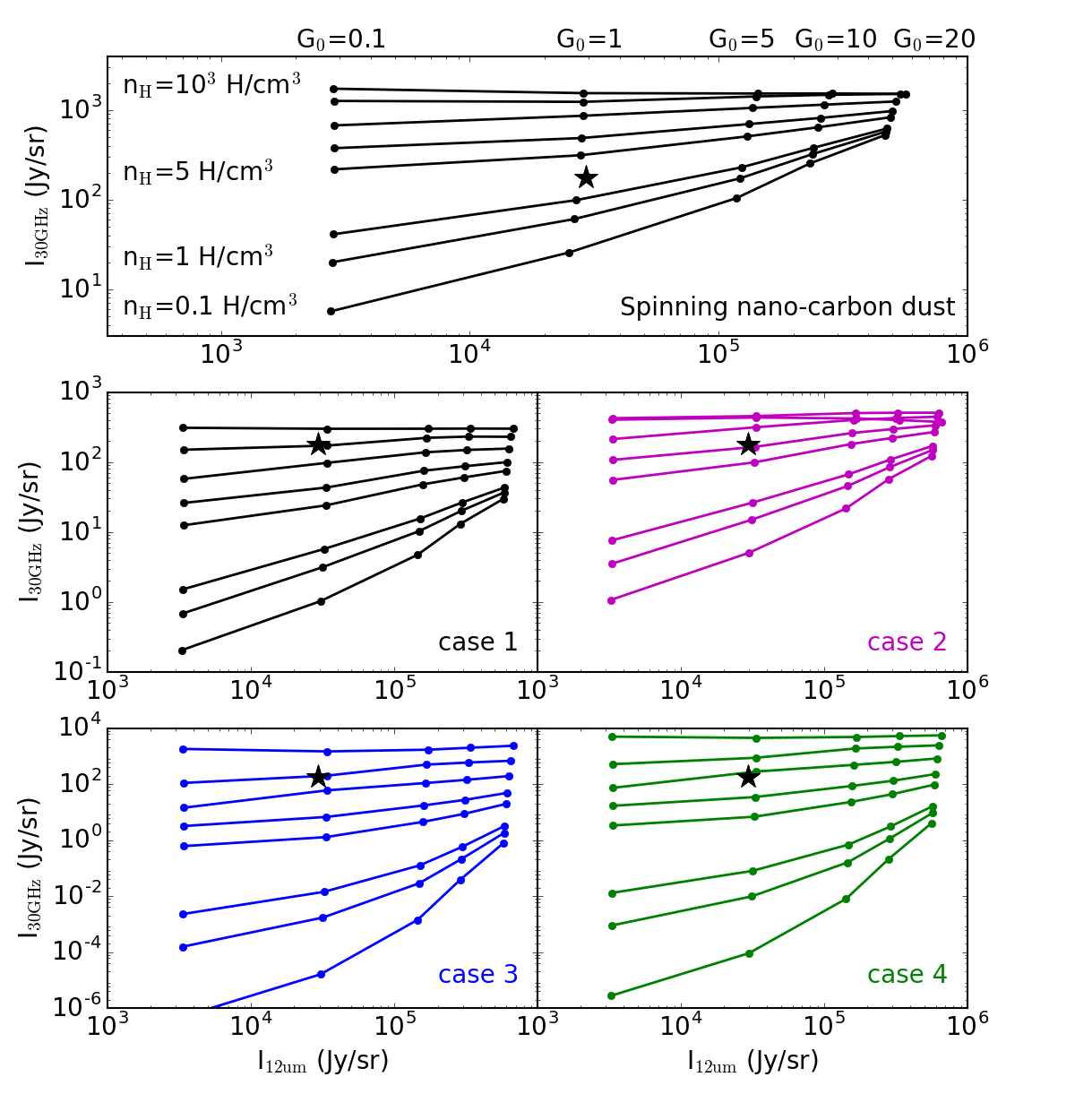}
\end{tabular}}
\caption{Thermal and spinning dust modelling results. Left: Dust SEDs for media illuminated by the ISRF with $G_0 = 1$. The dashed and solid lines show media with gas local densities of 30~H/cm$^3$ and 0.1~H/cm$^3$, respectively. The black star shows the average AME emissivity at 30~GHz for the bulk of pixels included in the 50\% contour in Fig.~\ref{figure_AME_vs_AME_12}, to avoid pixels with the highest and lowest densities and radiation fields. The upper panel shows spinning nano-carbon grains, and the bottom panel presents spinning nano-silicates with case 1 in black, case 2 in magenta, case 3 in blue, and case 4 in green; see Sect.~\ref{dust_properties} for details. Right: Spinning nano-dust emission at 30~GHz vs. thermal nano-dust emission at 12~$\mu$m for $N_{\rm HI} = 10^{20}$~H/cm$^2$. Each panel shows the results obtained for media with densities $n_{\rm H} = 0.1, 0.5, 1, 5, 10, 30, 100,$ and 1\,000~H/cm$^3$ from bottom to top. The radiation fields scaled by $G_0 = 0.1, 1, 5, 10,$ and 20 are shown from left to right. The upper panel displays spinning nano-carbons. The lower panels show the four cases of spinning nano-silicates described in Sect.~\ref{dust_properties}: case 1 in black, case 2 in magenta, case 3 in blue, and case 4 in green. The black star shows the average AME emissivity at 30~GHz vs. the average emissivity at 12~$\mu$m, both scaled to $N_{\rm H} = 10^{20}$~H/cm$^2$.}
\label{figure_modelling_results} 
\end{figure*}

\subsection{Resulting spectral energy distributions}
\label{modelling_results}

We show two sets of modelling results in Fig.~\ref{figure_modelling_results}. Firstly, the left panel shows the emission spectra of media with $G_0 = 1$ and local gas densities of $n_{\rm H} = 0.1$ and 30~H/cm$^3$ for spinning nano-carbons (top left) and spinning nano-silicates (bottom left). Secondly, the right panels present the predicted correlations between the mid-IR thermal emission in the IRAS 12~$\mu$m filter and the spinning dust emission at 30~GHz. These results highlight the spinning-dust intensity dispersion at 30~GHz as a function of $G_0$ and of the local gas density $n_{\rm H}$. Comparing lines of sight with the same column density that are dominated by low ($n_{\rm H} \sim 0. 1$~H/cm$^3$), moderate ($n_{\rm H} \sim 30$~H/cm$^3$), or high densities ($n_{\rm H} \sim 1\,000$~H/cm$^3$), the latter being typical of translucent clouds, we expect for $G_0 = 1$ spinning-dust 30~GHz intensity variations of about two orders of magnitude for nano-carbons, while for nano-silicates, with $m = 0.3$ and 1~D, the intensity varies by two and six orders of magnitude, respectively. The spinning nano-carbon models appear to be spread rather evenly about the observational data point, while the spinning nano-silicate models tend to lie below this point. We note that the 12~$\mu$m intensity remains almost constant for all of the considered models. This implies that the poor correlation of mid-IR emission to 30~GHz AME that is predicted from the spinning nano-carbon models must be even lower in the spinning nano-silicate model. This correlation does not improve when instead of the 12~$\mu$m thermal emission, we consider the emission at 100~$\mu$m. Therefore, the quality of the 30~GHz versus 12 or 100~$\mu$m correlation alone cannot be used to exclude any particular carrier. Then, if only the abundance of nano-carbon dust is allowed to vary (for a constant size distribution and constant environment properties), the spinning nano-carbon dust models predict a positive correlation between AME$_{\rm 30GHz}$/N$_{\rm HI}$ and $I_{12\mu{\rm m}}/R$, both proportional to the nano-carbon dust abundance. No such correlation is present in the observations \citep[see e.g.][]{Hensley2016}. This lack of correlation is a strong constraint on the models, and we further explore it in Sect.~\ref{models_vs_observations}. Spinning nano-silicate models predict a weaker correlation because the nano-silicates account for only $\sim 18\%$ of the total emission in the IRAS 12~$\mu$m band for $G_0 = 1$.

\begin{figure*}[!t]
\centerline{\begin{tabular}{cc}
\includegraphics[width=0.5\textwidth]{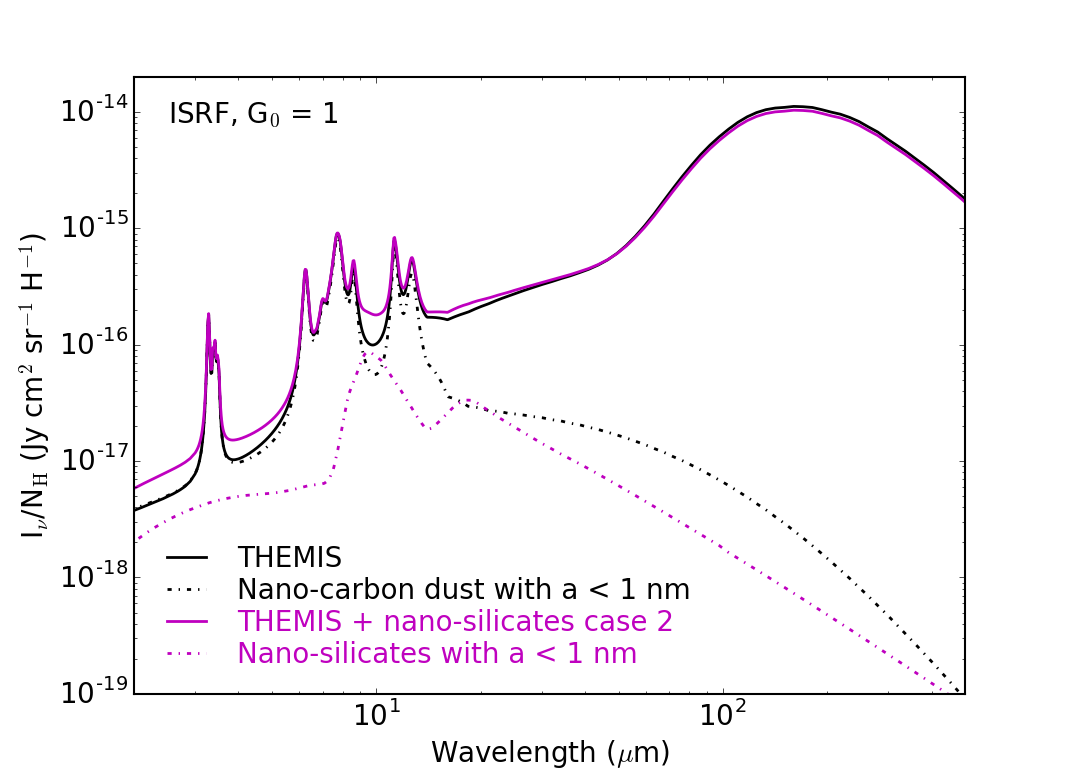} & \includegraphics[width=0.5\textwidth]{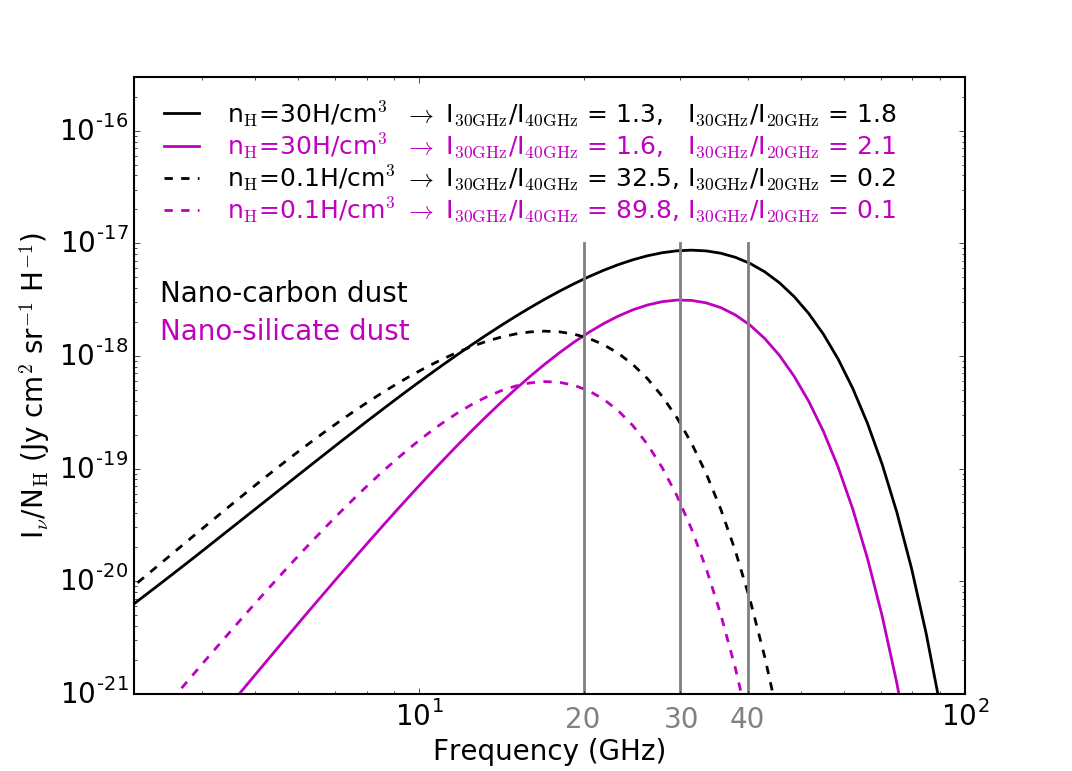}
\end{tabular}}
\caption{Comparison of mid-IR to microwave SEDs. Left: SED of thermal dust emission for dust models without (black lines) and with (magenta lines) nano-silicate dust for $G_0 = 1$. In both cases, the dash-dotted lines show the contribution of the smallest nano-grains with $a < 1$~nm that cause the spinning-dust emission. The nano-silicates are those of case 2, which are the closest in size distribution and electric dipole moment to the THEMIS nano-carbon dust. Right: Corresponding spinning-dust emission for media with $n_{\rm H} = 0.1$ and 30~H/cm$^3$as solid and dashed lines, respectively.}
\label{figure_sed_spin} 
\end{figure*}

Based on the results presented in Figs.~\ref{figure_modelling_results} and \ref{figure_sed_spin}, we make two remarks concerning the comparison of the models with observations. Firstly, to achieve the same 30~GHz intensity, higher local densities are required for spinning nano-silicates than for nano-carbons. To obtain the same intensity as spinning nano-carbons in a medium with $n_{\rm H} = 5$~H/cm$^3$ at $G_0 = 1$, for example, the nano-silicate model requires densities of about 100 to 1\,000~H/cm$^3$. This difference is mainly explained by the difference in the contribution of the IR emission to the rotational excitation of the sub-nanometre grains responsible for the SED of the spinning dust (left panel in Fig.~\ref{figure_sed_spin}). In their Eqs.~ 95 and 106, \citet{Silsbee2011} indeed showed that the IR excitation and damping rates are proportional to $\int_0^\infty I_{\nu}/\nu^2 \, d\nu$ and $\int_0^\infty I_{\nu}/\nu \, d\nu$, respectively. Thus, for the case $n_{\rm H} = 0.1$~H/cm$^3$ and $G_0 = 1$ presented in Fig.~\ref{figure_sed_spin} (dashed lines), the rotational excitation by IR emission is a factor 0.8 lower and the braking is a factor 1.5 higher for a 0.45~nm nano-silicate than for a nano-carbon of the same size. These discrepancies become even larger with increasing size (factors 0.6 and 7.5, respectively, for 0.6~nm nano-grains), which explains that the SEDs for nano-silicates are narrower than for nano-carbons on the low frequency side (right panel in Fig~\ref{figure_sed_spin}). These differences in the spinning SED intensity and width imply that if the gas distribution in the different interstellar medium phases along a given line of sight is known, it should be possible in principle to determine the types of grains that are the dominant AME carriers. Then, regardless of the gas density used to model the spinning dust emission, the peak frequency and the width of the spectra differ sufficiently between spinning nano-carbons and nano-silicates to result in significantly different I$_{\rm 30GHz}$/I$_{\rm 40GHz}$ and I$_{\rm 30GHz}$/I$_{\rm 20GHz}$ intensity ratios (see the example presented in the right panel of Fig.\ref{figure_sed_spin}). This indicates that several frequencies and not only one (i.e. 30~GHz) are required to characterise AME. Correlating the 30~GHz AME with the 20 and 40~GHz AME is therefore more discriminating than using the mid- to far-IR emission to characterise the AME carriers.

\section{CNM versus WNM fraction in the diffuse neutral interstellar medium}
\label{CNM_vs_WNM}

Unlike thermal dust emission, spinning dust emission depends not only on the radiation field, but also on the gas density, as we showed in Sect. 3.3. The local gas properties impact the total intensity of the rotational spectrum, its peak frequency, and its width. This implies that to effectively model the AME and its correlation with other observed quantities, a realistic estimate of the gas density distribution is essential.

The diffuse neutral medium on which we focus can be described as a two-phase medium \citep[e.g.][]{Field1969, McKee1977, Wolfire2003}. At thermal equilibrium, the principal heating and cooling processes acting on the interstellar gas allow two thermally stable HI phases that are distinguished by their density and temperature: a cold neutral medium (CNM) with a low kinetic temperature $T \sim 30 - 200$~K and medium density $n_{\rm H} \sim 5 - 120$~H/cm$^3$ , and a warm neutral medium (WNM) with a high temperature $T \sim 4\,100 - 8\,800$~K and low density $n_{\rm H} \sim 0.03 - 1.3$~H/cm$^3$. These two phases have been confirmed observationally \citep[e.g.][]{Liszt1993, Heiles2003, NGuyen2019, Murray2020}. However, it has been shown that a significant fraction of the WNM is cooler than expected ($T \sim 500 - 5\,000$~K) and thus falls into the thermally unstable regime. The exact proportion of this unstable HI is still debated \citep[e.g.][]{Heiles2003, Murray2018}, however.

Observations show that the CNM fraction, $f_{\rm CNM} = N_{\rm HI}^{\rm CNM} / (N_{\rm HI}^{\rm CNM} + N_{\rm HI}^{\rm WNM})$, increases with the total column density of HI and in the surroundings of molecular clouds as well \citep[e.g.][]{Stanimirovic2014}. For very diffuse high-latitude sightlines, with HI column densities ranging from $3\times 10^{16}$ to a few $10^{21}$~cm$^2$, \citet{Murray2015} found that $f_{\rm CNM}$ extends from less than 0.1 to 0.51, with a median value of 0.20. This agrees with the median $f_{\rm CNM} = 0.23$ measured by \citet{Heiles2003} for sightlines with $|b| > 10^\circ$ and $N_{\rm HI}$ from a few $10^{20}$ to a few $10^{21}$~H/cm$^2$. \citet{Stanimirovic2014} further showed that for column densities from $\sim 10^{18}$ to a few $10^{21}$~H/cm$^2$, the HI column density of the WNM is relatively uniform, whereas at $N_{\rm HI} \sim 10^{21}$~H/cm$^2$ , the CNM fraction increases from almost negligible to $\sim 40\%$ around Perseus, with a median value of 33\%. Regions such as Perseus are representative of the areas we selected for our study (Fig.~\ref{figure_mask}), and it is indeed included in the mask defined in Sect.~\ref{masks}, as are the Taurus and Gemini regions, for which \citet{NGuyen2019} found $f_{\rm CNM} = 37$\% and 16\%, respectively. Similar results were found in the California, Rosette, MonOB1, and NGC2264 regions \citep{NGuyen2019}, which are also included in our mask.

Of the sight lines included in our study, 92\% have $5\times 10^{20} \leqslant N_{\rm HI} \leqslant 5\times 10^{21}$~H/cm$^2$ and are above the Galactic Plane. We therefore make the reasonable assumption below that in order to model AME by spinning nano-dust emission, a model is acceptable only if it uses a mixture of CNM and WNM with $0.05 \leqslant f_{\rm CNM} \leqslant 0.4$ and if it matches the AME and mid-IR brightness for HI column densities matching the distribution shown in Fig.~\ref{figure_NHI_distribution}.

\section{Modelling results}
\label{models_vs_observations}

\begin{figure*}[!t]
\centerline{\begin{tabular}{c}
\includegraphics[width=1\textwidth]{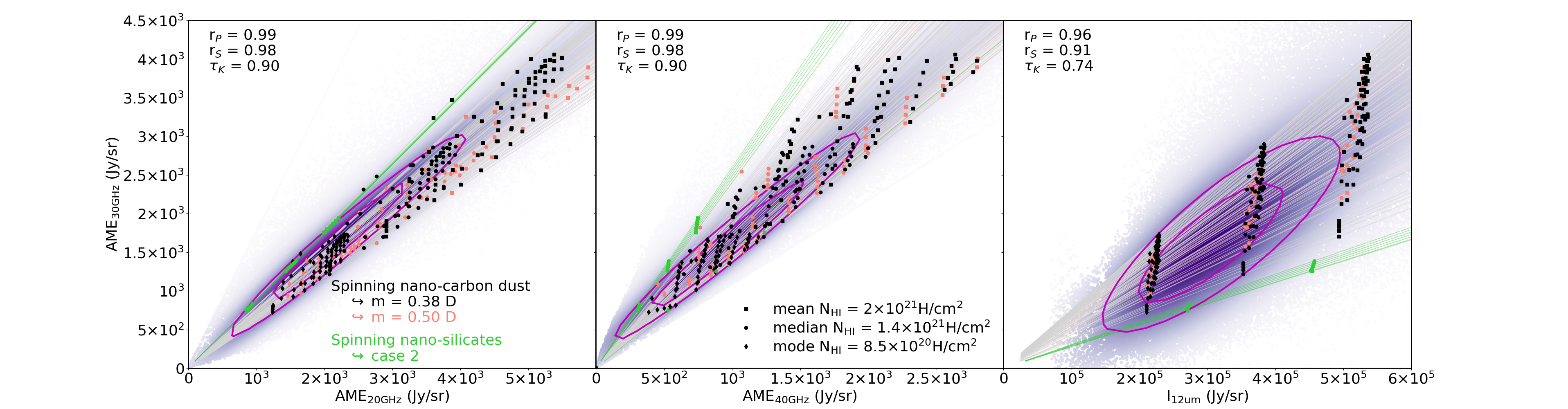} \\
\includegraphics[width=1\textwidth]{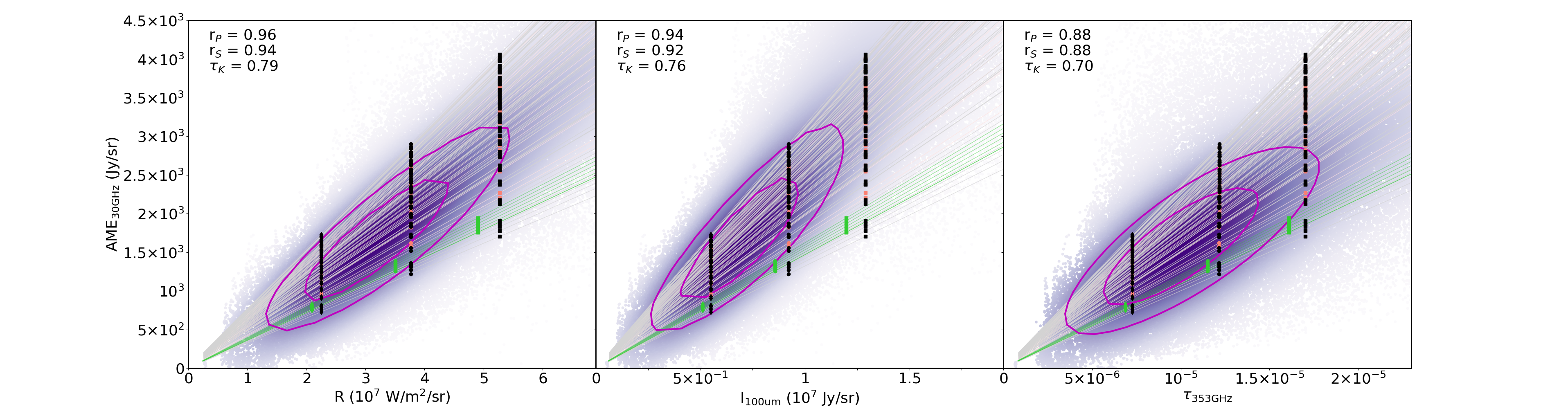}
\end{tabular}}
\caption{Comparison of observed and modelled correlations. Top from left to right: observations are plotted as density of points maps of AME$_{\rm 30~GHz}$ vs. AME$_{\rm 20~GHz}$, vs. AME$_{\rm 40~GHz}$ , and vs. dust emission in the 12~$\mu$m IRAS filter. Bottom from left to right: Observations are plotted as density of points maps of AME$_{\rm 30~GHz}$ vs. radiance, vs. dust emission in the 100~$\mu$m IRAS filter, and vs. optical depth at 353~GHz. For all panels, we overplot the contours enclosing 50\% of the pixels (internal contour) and 75\% of the pixels (external contour). The models fitting the observations are shown with black and pink symbols for spinning nano-carbon dust, $m = 0.38$ and 0.5~D, respectively, and green symbols for spinning nano-silicate dust (case 2; see Sect.~\ref{models_vs_observations} for details). Squares, circles, and diamonds show models scaled to the mean ($N_{\rm HI} = 2\times10^{21}$~H/cm$^2$), the median ($N_{\rm HI} = 1.4\times10^{21}$~H/cm$^2$), and the mode HI column density ($N_{\rm HI} = 8.5\times10^{20}$~H/cm$^2$), respectively. The light grey, pink, and green lines show the same models for HI column densities extending over the entire obervational range $10^{20} \leqslant N_{\rm HI} \leqslant 1.4 \times 10^{22}$~H/cm$^2$. The various model points show spinning-dust models for different $f_{\rm CNM}$ values (see Sect.~\ref{nano-carbons} for details). The $r_P$, $r_S$ , and $\tau_K$ give the Pearson correlation coefficient, the Spearman rank-order correlation coefficient, and the Kendall's tau, respectively.}
\label{figure_AME_vs_AME_12} 
\end{figure*}

\begin{figure}[!t]
\centerline{\includegraphics[width=0.5\textwidth]{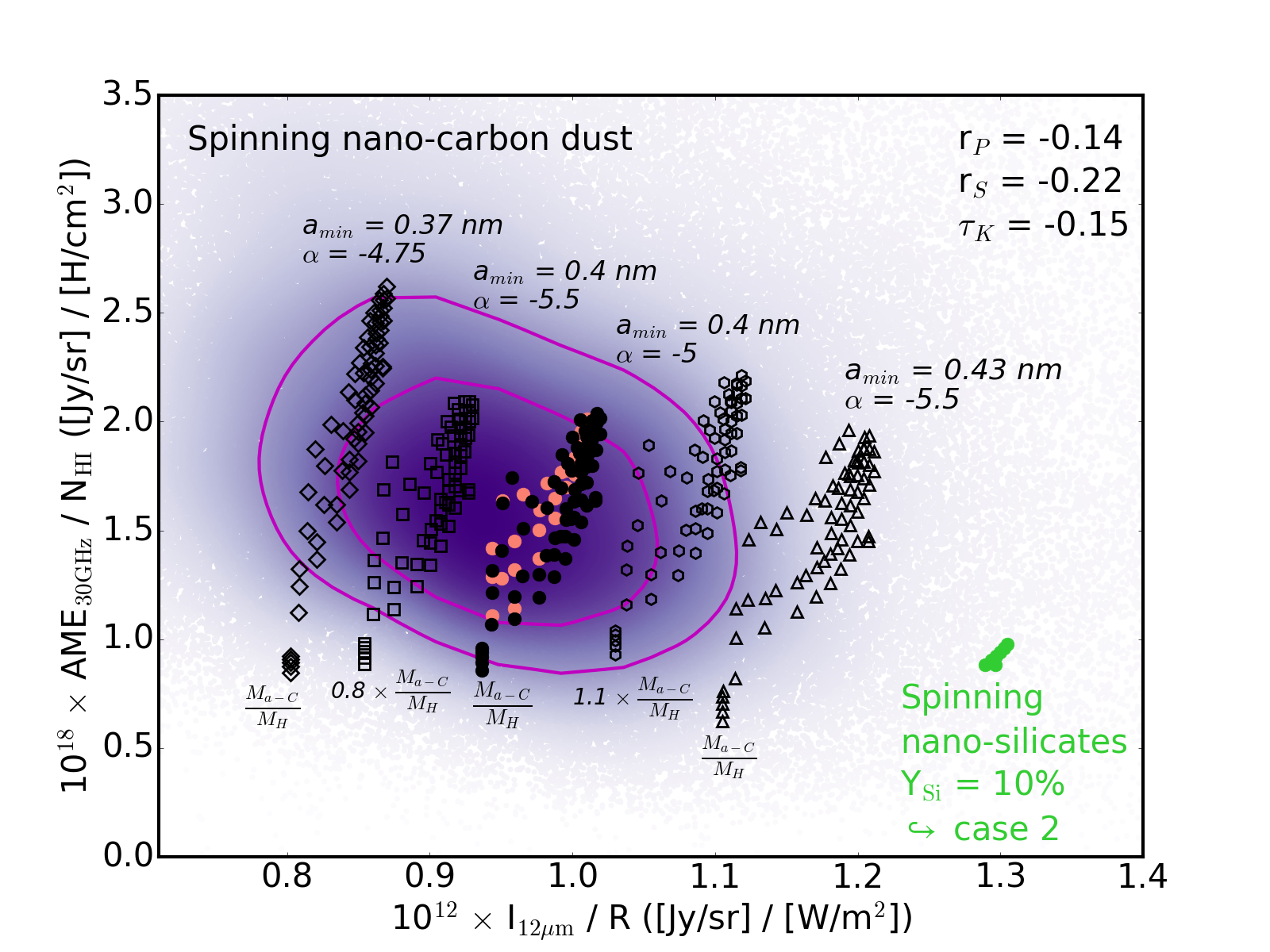}}
\caption{Density of the points map of AME$_{\rm 30~GHz}$/N$_\mathrm{HI}$ vs. $I_{12\mu{\rm m}}/R$. We overplot the contours enclosing 50\% of the pixels (internal contour) and 75\% of the pixels (external contour). The $r_P$, $r_S$ , and $\tau_K$ give the corresponding Pearson correlation coefficient, the Spearman rank-order correlation coefficient, and the Kendall's tau, respectively. The same models as in Fig.~\ref{figure_AME_vs_AME_12} are shown with black and pink circles for spinning nano-carbon dust, $m = 0.38$ and 0.5~D, respectively, and green circles for spinning nano-silicate dust (case 2), each symbol showing a different value of $f_{\rm CNM}$ (see Sects.~\ref{nano-carbons} and \ref{nano-silicates} for details). For spinning nano-carbon dust, we show models with the same $f_{\rm CNM}$ but with variations in the size distribution and dust abundance. The standard THEMIS size distribution with $a_{min} = 0.4$~nm and $\alpha = -5$, where $\alpha$ is the a-C nano-particle power-law exponent, is shown for $m = 0.38$~D (black circles). We show this same model, but with an abundance increase of 10\% (hexagons). The squares show $a_{min} = 0.4$~nm, $\alpha = -5.5,$ and an abundance decrease by 20\%. The triangles show $a_{min} = 0.43$~nm and $\alpha = -5.5$, and the diamonds show $a_{min} = 0.37$~nm and $\alpha = -4.75$ with the standard model abundances.}
\label{figure_AME_12_R} 
\end{figure}

To test the different models, we considered CNM/WNM mixtures with $f_{\rm CNM} = 0$ to 0.5 in steps of 0.05 illuminated by the \citet{Mathis1983} radiation field with $G_0 = 1$. According to \citet{Fanciullo2015}, the typical $G_0$ variations range from $\sim 0.8$ to 1.4. The HI column density was set equal to the median, the mean, or the mode of the $N_{\rm HI}$ distribution (see Fig.~\ref{figure_NHI_distribution}). A model was considered acceptable when inside the contours that encompasse 75\% of the pixels simultaneously for all three ratios of AME$_{\rm 30~GHz}$ versus AME$_{\rm 20~GHz}$, AME$_{\rm 30~GHz}$ versus AME$_{\rm 40~GHz}$ , and AME$_{\rm 30~GHz}$ versus dust emission in the IRAS 12~$\mu$m filter for at least one of the tested column densities. All models were run for both nano-carbons and nano-silicates with the properties described in Sect.~\ref{dust_properties}. Figures~\ref{figure_AME_vs_AME_12} and \ref{figure_AME_12_R} show the results.

\subsection{Digression from observation correlations}
\label{digression}

Before we describe the details of the modelling, we explored the level of correlation between the different data sets (Fig.~\ref{figure_AME_vs_AME_12}). Thanks to the new treatment applied to the IRAS data (see Appendix~\ref{IRIS}), which excludes the noisiest areas (see Sect.~\ref{masks}), the strength of the linear correlation of the AME for an angular resolution of 1$^\degree$ is as good with the emission at 12~$\mu$m as with the emission at 100~$\mu$m: the Pearson correlation coefficients are $r_P = 0.96$ and 0.94, respectively. As expected, matter correlates with matter, which shows that our results will not be limited by data problems, but simply by our level of understanding of the dust. Figure~\ref{figure_AME_vs_AME_12} shows that the data dispersion remains relatively high, which needs to be explained either by different environments (our mask includes 24\% of the sky) or by variations in the nano-grain properties (nature, size distribution, abundance, and electric dipole moment). If suitable matches between the models and the data are not possible, then the spinning nano-grain hypothesis has to be discarded.

\subsection{Spinning nano-carbon dust}
\label{nano-carbons}

The model of spinning nano-carbon grains is able to reproduce the observed trends in Fig.~\ref{figure_AME_vs_AME_12} (black symbols), but these results do not allow us to constrain the density of the WNM because all of the tested values, $0.05 \leq n_{\rm H}^{\rm WNM} \leq 1$~H/cm$^3$, yield solutions for pixels within both the 75\% and 50\% contours. However, we do note two interesting results. Firstly, the CNM density depends on the WNM density, and a denser WNM implies a denser CNM. For example, for models with a $n_{\rm H}^{\rm WNM} = 0.1$~H/cm$^3$, the CNM density must be between 20 and 40~H/cm$^3$. If $n_{\rm H}^{\rm WNM} = 0.6$~H/cm$^3$, however, $n_{\rm H}^{\rm CNM}$ must be between 30 and 60~H/cm$^3$, and if $n_{\rm H}^{\rm WNM} = 1$~H/cm$^3$, then $n_{\rm H}^{\rm CNM}$ must be between 40 and 100~H/cm$^3$. Secondly, all viable models require $0.1 \leq f_{\rm CNM} \leq 0.2$, which is in agreement with the observational studies cited in Sect.~\ref{CNM_vs_WNM}. This also agrees with the results of \citet{Ysard2010b}, who used the IRIS map at 12~$\mu$m and the 23~GHz AME map derived by \citet{MAMD2008} from WMAP data to study 27 regions of a few square degrees, which are all included in our mask. With a spinning PAH model, they obtained $f_{\rm CNM} \sim 10\%$ for all regions and $m \sim 0.4$~D for most of them. This is a good indication that models using carbonaceous nano-grains give comparable results regardless of their exact nature.

The grain electric dipole moment is probably the most uncertain dust parameter in our modelling. We therefore also show the results for the same nano-carbon dust, but with a higher dipole moment of $m = 0.5$~D (pink symbols in Fig.~\ref{figure_AME_vs_AME_12}). The results for spinning nano-carbon grains still hold, but the required density of the WNM for the models that fit cannot exceed 0.7~H/cm$^3$. With a higher electric dipole moment, the spectrum of the spinning nano-carbon shifts to the lower frequency side that is dominated by the WNM contribution. To avoid overestimating the intensity at 20~GHz, the rotational excitation by the gas in the WNM must be lower. This requires a lower gas density.

For consistency, we also show the AME$_{\rm 30GHz}$ correlation with radiance (proportional to the radiation field and the colum density), the intensity in the IRAS 100~$\mu$m band (emission from large grains at thermal equilibrium), and the dust opacity at 353~GHz. All models that are in accordance with the AME and the mid-IR emission also agree with these far-IR/submm emission (derived) observations. This illustrates the consistency of THEMIS at all wavelengths.

Figure \ref{figure_AME_12_R} shows AME$_{\rm 30GHz}$/N$_{\rm HI}$ as a function of $I_{12\mu{\rm m}}/R$. When we assume that there is no nano-silicate dust in the diffuse interstellar medium, $I_{12\mu{\rm m}}/R$ is proportional to the nano-carbon dust abundance. According to \citet{Fanciullo2015}, the typical $G_0$ variations range from $\sim 0.8$ to 1.4 in the diffuse interstellar medium, which leads to variations in AME$_{\rm 30GHz}$/N$_{\rm HI}$ of $\sim -4$ to +8\% for the CNM/WNM mixtures required by the viable spinning-dust models shown in Fig.~\ref{figure_AME_vs_AME_12}. The models that best account for the observations presented so far also agree with the correlations in the ratios (full black and pink dots in Fig.~\ref{figure_AME_12_R} for $m = 0.38$ and 0.5~D, respectively). If it is indeed due to spinning dust, the AME$_{\rm 30GHz}$/N$_{\rm HI}$ ratio should be roughly proportional to the nano-carbon dust abundance. There is no obvious correlation in the data with a Pearson correlation coefficient of -0.14. On the one hand, this might be an illustration of the fact that the excitation of the rotation of nano-grains depends on interactions with the gas and also on the UV radiation field, whereas the mid-IR emission only depends on the latter. On the other hand, the observed dispersion could be the result of small variations in the nano-grain size distribution. The nano-carbon grain dust population, whether PAHs or amorphous hydrocarbons is most strongly affected by stellar photo-processing and shock waves, which both lead to variations in their size distribution and abundance \citep[e.g.][]{Bocchio2014, Murga2019, Murga2020, Joblin2020, Schirmer2020}. The observed dispersion in the correlations of the ratios can be explained by small variations in the size distribution and abundance (black symbols in Fig.~\ref{figure_AME_12_R}). Photo-processing studies show that a decrease in the nano-grain abundance is also associated with a change in the size distribution. For example, \citet{Schirmer2020} showed that in the Horsehead nebula, the slope of the nano-carbon dust size distribution steepens from $n(a) \propto a^{-5}$ to $n(a) \propto a^{\rm -5.5~to~-7.5}$. For the least extreme case, $n(a) \propto a^{-5.5}$ and a 20\% decrease in the grain abundance, we obtain the squares in Fig.~\ref{figure_AME_12_R}. Our results also show that a slight increase in the nano-carbon dust abundance (hexagons), a small increase in the minimum size coupled to a steeper size distribution slope (triangles), or a shallower size distribution coupled to a smaller minimum size (diamonds) can explain the observed dispersion. The minimum size and slope of the size distribution are indeed degenerate because they both control the quantity of the smallest nano-grains, which cause the AME. These models are also consistent with the correlations presented in Fig.~\ref{figure_AME_vs_AME_12}.

\subsection{Spinning nano-silicate dust}
\label{nano-silicates}

For the spinning nano-silicates with the highest electric dipole moment (cases 3 and 4, $m = 1$~D), it is impossible to simultaneously fit the ratios of AME$_{\rm 30~GHz}$ versus AME$_{\rm 20~GHz}$ and AME$_{\rm 30~GHz}$ versus AME$_{\rm 40~GHz}$. This electric dipole moment is indeed high enough to significantly slow the rotation speed down and thus leads to spinning emission that peaks at lower frequencies than observed. This implies that even if it is possible to reproduce the intensity level at a single given frequency, for instance at 30~GHz, the spectral shape will be incorrect and accordingly, the required intensities at the two other frequencies cannot be retrieved. We note that the quantum mechanical calculations performed by \citet{Macia2020} and \citet{Marinoso2021} show that Mg-rich nano-silicates will always have $m > 1$~D, even when they are covered by an ice layer. However, high dipole moments like this for nano-silicates appear to be inconsistent with the observations \citep[see also][]{Hoang2016, Hensley2017}. 

Only nano-silicates with a lower dipole moment and a size distribution centred on the smallest grains result in a spinning emission within the 75\% contour (case 2: $m = 0.3$~D and $\sigma = 0.1$, green symbols in Fig.~\ref{figure_AME_vs_AME_12}). However, this works for only six CNM/WNM mixtures, all having $f_{\rm CNM} = 0.3$ and a CNM density $n_{\rm H} = 20$~H/cm$^3$, with the WNM density varying from 0.5 to 1~H/cm$^3$. Only so few models fall within the 75\% contour because the intensity in the IRAS 12~$\mu$m band is too high. This is more obvious when we plot the AME$_{\rm 30GHz}$/N$_{\rm HI}$ ratio over $I_{12\mu{\rm m}}/R,$ for which the nano-silicates are within the observed dispersion, but not within the 75\% contour (green dots in Fig.~\ref{figure_AME_12_R}). This leads us to the following conclusion: if $Y_{\rm Si} = 10\%$, then nano-silicates can only marginally contribute to the AME, regardless of their properties.

The only way to reduce the nano-silicate mid-IR emission is to decrease $Y_{\rm Si}$. However, for $G_0 = 1$, the nano-silicates only account for about $18\%$ of the total intensity because the IRAS 12~$\mu$m filter is very wide ($\sim 7~\mu$m). They do, however, account for 45\% of the total intensity at the peak frequency of the relatively narrow silicate spectral feature at 9.8~$\mu$m. As a result, halving $Y_{\rm Si}$, only decreases $I_{12\mu{\rm m}}$ by about 9\%, while the 30~GHz spinning emission is decreased by 50\%. We therefore come to the same conclusion, that nano-silicates can, at best, only contribute marginally to the observed AME.

\citet{Hensley2017} suggested that uncertainties in the grain equilibrium charge distribution could lead to an uncertainty in the nano-silicate spinning SED, in particular if the number of negatively charged grains is underestimated. These are indeed more rotationally excited. As discussed in \citet{Weingartner2001}, the grain charge distribution and the number of negatively charged grains depends on the bulk work function $W,$ which we take to be 4.97~eV for silicates, following \citet[][and references therein]{Kimura2016} instead of 8~eV, which is the standard value often adopted for all grains, whether they are based on carbon or silicate \citep{Draine1978}. To test the influence of this choice, we ran calculations with $W = 8$~eV. The same fitting procedure changes our results very little: only case 2 nano-silicates give a few models within the 75\% contour for the same CNM and WNM gas densities. The only difference is that $f_{\rm CNM}$ is decreased to 0.25. This is a consequence of the fact that negatively charged grains are not abundant in diffuse gas that is pervaded by UV.

Our conclusions might be biased because THEMIS comprises amorphous aromatic-rich nano-carbon grains rather than so-called `astro-PAHs'. This is not the case. For the PAHs of the \citet{Compiegne2011} model, the difference in the IRAS 12~$\mu$m band is only $\sim 3\%$ in favour of THEMIS. The difference might come from the adopted silicate optical properties, but this is again not the case. When we compare the \citet{Draine2007} model, which contains both PAHs and nano-silicates in more or less the same proportions as our case 2, at $G_0$ = 1, the difference in the IRAS 12~$\mu$m band is $\sim 5\%$, which is again in favour of THEMIS. This means that for the diffuse interstellar medium sky area (Fig.~\ref{figure_mask}) we considered, all models that attempt to explain the AME, at several frequencies, with nano-silicates alone have the same problem, whether they use PAHs or amorphous nano-carbon grains. \citet{Hoang2016} did not have this problem because they did not consider the mid-IR emission. They tested their model against the UV extinction, polarisation of starlight, and the polarisation of the AME. They were able to place constraints on both $Y_{\rm Si} \leqslant 10\%$ and $m \geqslant 0.4$~D (see their Fig.~8). With regard to the results of \citet{Hensley2016}, $4 \leqslant Y_{\rm Si} \leqslant 15\%$ and $0.3 \leqslant m \leqslant 1$~D, the difference can probably be reduced to a single aspect: the normalisation of their spectra. Their mid-IR spectrum is that of the translucent cloud DCld 300.2-16.9. According to \citet[][and references therein]{Ingalls2011}, DCld 300.2-16.9 is a CNM cloud that is illuminated by the ISRF with $G_0 = 1$, with a column density of $\sim 4 \times 10^{21}$~H/cm$^2$, that was affected by supernova explosions from the Sco-Cen OB association about (2-6)$\times 10^5$~years ago. This spectrum was scaled to the HI-correlated dust emission in the DIRBE bands as in \citet{Dwek1997}: $I_{12\mu{\rm m}} = 3.16 \times 10^{-16}$~Jy cm$^2$/sr/H and was then associated with the median AME Planck COMMANDER spectrum normalised to $3\times 10^{-18}$~Jy cm$^2$/sr/H at 30~GHz. This is far from the the bulk of the pixels of the intermediate-latitude regions that we studied here. 

\begin{figure}[!t]
\centerline{\includegraphics[width=0.5\textwidth]{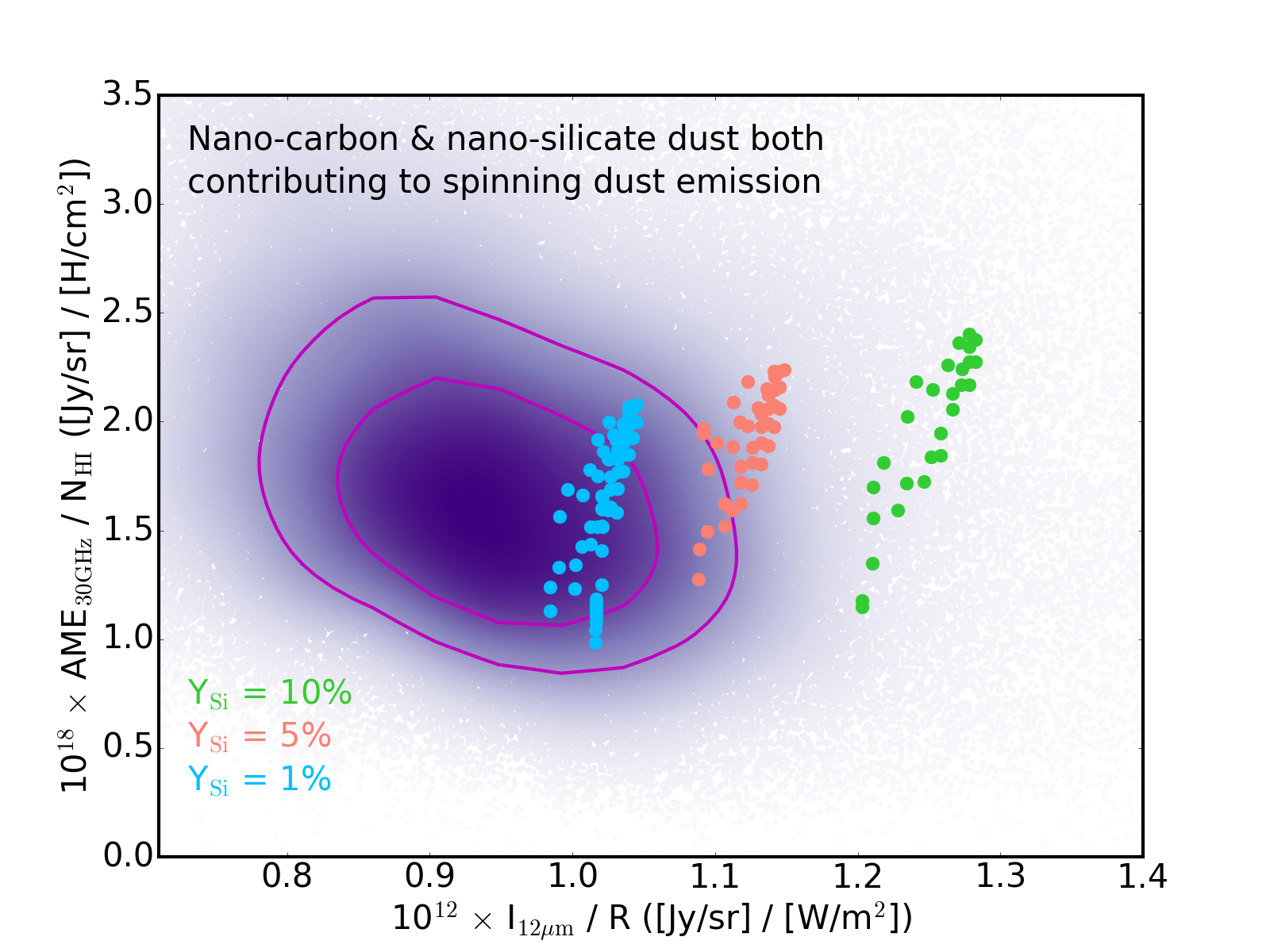}}
\caption{AME$_{\rm 30~GHz}$/N$_\mathrm{HI}$ vs. $I_{12\mu{\rm m}}/R$ as in Fig.~\ref{figure_AME_12_R}. The green dots show best fitting models in which both the nano-carbon dust and the nano-silicate dust produce spinning emission (case 2, $Y_{\rm Si} = 10\%$, $m = 0.3$~D, $\sigma = 0.1$), with each point showing a different $f_{\rm CNM}$ value. Model calculations for lower abundances of nano-silicates are shown: $Y_{\rm Si} = 5\%$ (pink dots) and $Y_{\rm Si} = 1\%$ (blue dots).}
\label{figure_both_spinning} 
\end{figure}

\subsection{Combined nano-carbon and nano-silicate dust emission}
\label{both_spinning}

If nano-carbon and nano-silicate dust co-exist in the diffuse interstellar medium, it is reasonable to assume that both will produce spinning dust emission. For carbonaceous nano-dust, we have clear and abundant observational evidence for their existence in non-negligible quantities in the interstellar medium, while the case is not so clear-cut for nano-silicates. In order to be comprehensive, we therefore repeated the fitting procedure with the assumption that both populations contribute to the AME. We did this for several values of $Y_{\rm Si}$ ($= 10$, 5 and 1\%) for case 2 nano-silicates ($m = 0.3$~D and $\sigma = 0.1$). All of these nano-silicate abundances led to acceptable CNM/WNM mixtures, and we show the corresponding model results in Fig.~\ref{figure_both_spinning}. The default value $Y_{\rm Si} = 10\%$ is within the dispersion, but does not fall within the contours surrounding the bulk of the pixels. If the abundance of nano-silicates is halved, then the models are at the edge of the 75\% contour. When the abundance is further lowered to $Y_{\rm Si} = 1\%$, the models fall within the 50\% contour, but remain on the high $I_{12 \mu{\rm m}}/R$ side, and nano-silicates account only for 3 - 5\% of the AME at 30~GHz, depending on the CNM/WNM mixture. We do not show cases 1, 3, and 4 because the discrepancy comes mainly from the $I_{12 \mu{\rm m}}/R$ ratio. Cases 2 and 4 have the exact same $I_{12 \mu{\rm m}}/R$, which does not depend on $m,$ but just on $\sigma$. Cases 1 and 3 are similar and have the same radiance (no changes in the larger grain populations), and the $I_{12 \mu{\rm m}}$ values vary by no more than 2\% when compared to cases 2 and 4.

Fig.~\ref{figure_both_spinning} indicates that (1) if both nano-grain populations co-exist in the diffuse interstellar medium, then the AME must come predominantly from nano-carbon dust emission, and (2) if both populations have non-zero permanent electric dipole moments, then the abundance of nano-silicates cannot exceed $\sim 1$\%. This upper limit agrees with the findings of \citet{Desert1986}, \citet{Macia2020}, and \citet{Marinoso2021}. All three studies derive an upper limit of $Y_{\rm Si} = 1\%$. This is lower than the upper limit of $\sim 10$\% reported by \citet{Li2001}. The difference most probably comes from the line of sight chosen by \citet{Li2001} to test their model ($l = 44\degree 20'$, $b = -0\degree 20'$), which is in the Galactic Plane and has a relatively high column density ($N_{\rm H} = 4.3 \times 10^{22}$~H/cm$^2$) and a radiation field with $G_0 = 2$. This line of sight is therefore not representative of the bulk of the diffuse interstellar medium lines of sight at intermediate latitudes.

\section{Conclusion}
\label{conclusion}

Using The Heterogeneous dust Evolution Model for Interstellar Solids \citep[THEMIS, ][]{Jones2017}, we investigated whether the nature of the grains at the origin of the AME measured by the Planck Collaboration \citep{COMMANDER2016} in the Galactic diffuse interstellar medium can be identified. Model grids were created with the DustEM numerical tool \citep{Compiegne2011} and were compared with observations and parameters derived from them, from the mid-IR to the microwave. Our results are listed below.

First, nano-carbon dust can explain all the observations for medium properties, in agreement with the latest findings about the separation of CNM and WNM in the diffuse interstellar medium. The dispersion in the observations can be accounted for with little variations in the dust size distribution, abundance, or electric dipole moment. Second, regardless of the properties and the abundance of the nano-silicate dust considered, spinning nano-silicates are formally excluded as the sole source of the AME. Third, the best agreement with observations is obtained when the emission of spinning nano-carbons \textit{\textup{alone}} is taken into account. The addition of a nano-silicate component takes the model beyond the maximum density area of the correlation plots. However, a marginal participation of these nano-silicates to the AME production cannot be excluded as long as their abundance does not exceed $Y_{\rm Si} \sim 1\%$.

Our results reconcile large- and small-scale studies. Careful investigations of the carrier nature and evolution through modelling and data correlations, have linked AME and spinning nano-carbonaceous dust in these studies, be they amorphous hydrocarbons or PAHs. The AME is thus very promising for tracing the evolution of carbonaceous grains because the spinning emission does not only depend on the grain properties, but also on the properties of the immediate local medium. This should allow testing the models of grain evolution in a strongly constraining way, whether they are due to UV photons or shocks.

\begin{acknowledgements}
We thank our anonymous referee whose careful reading and interesting comments helped to clarify and improve the paper. This work was supported by the Programme National PCMI of CNRS/INSU with INC/INP co-funded by CEA and CNES. Finally, thanks to Dan Pineau who helped me fix my office chair, an essential tool if ever there was one.
\end{acknowledgements}

\bibliography{biblio}

\begin{appendix}

\section{Removal of zodiacal residuals at 12~$\mu$m}
\label{IRIS}

\begin{figure}[!t]
\centerline{\begin{tabular}{c}
\includegraphics[width=0.36\textwidth]{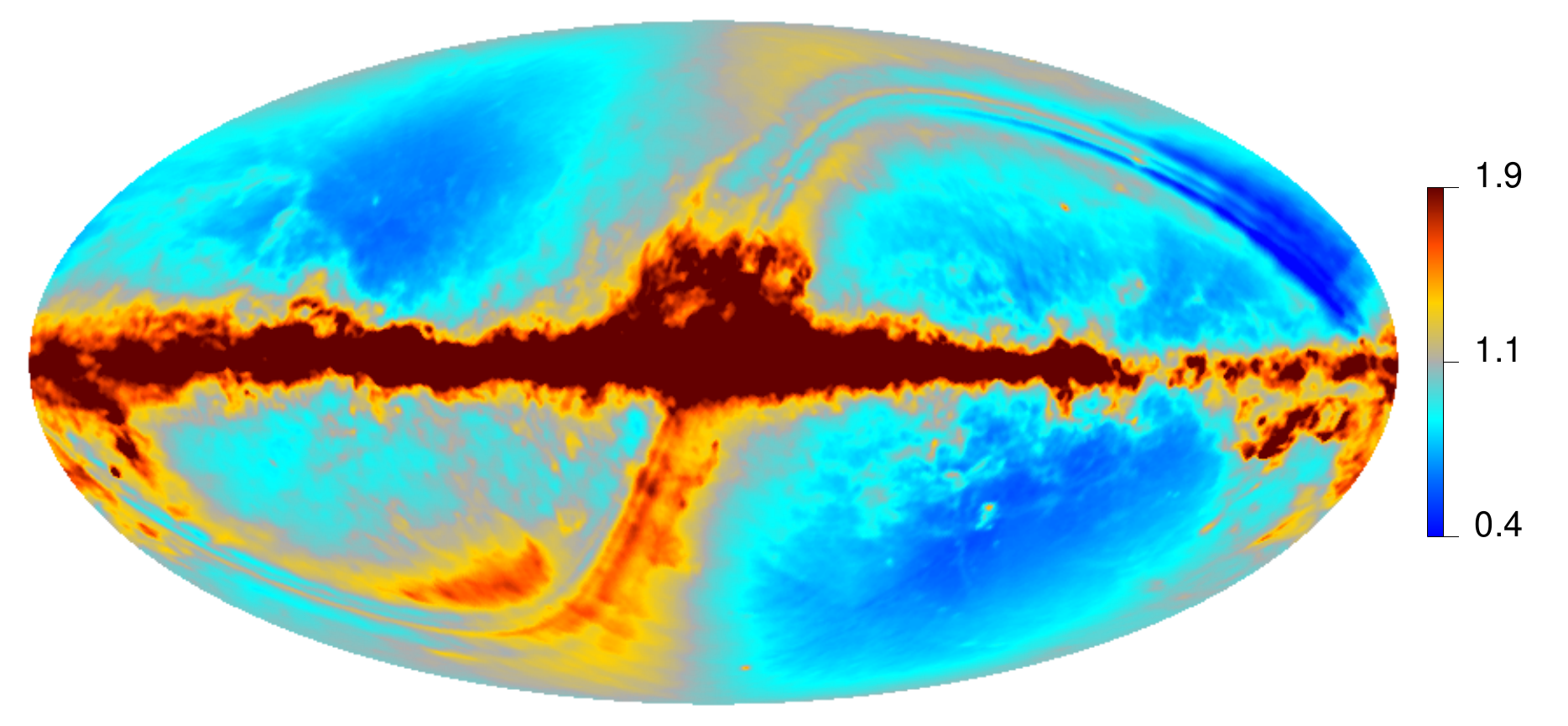} \\
\includegraphics[width=0.36\textwidth]{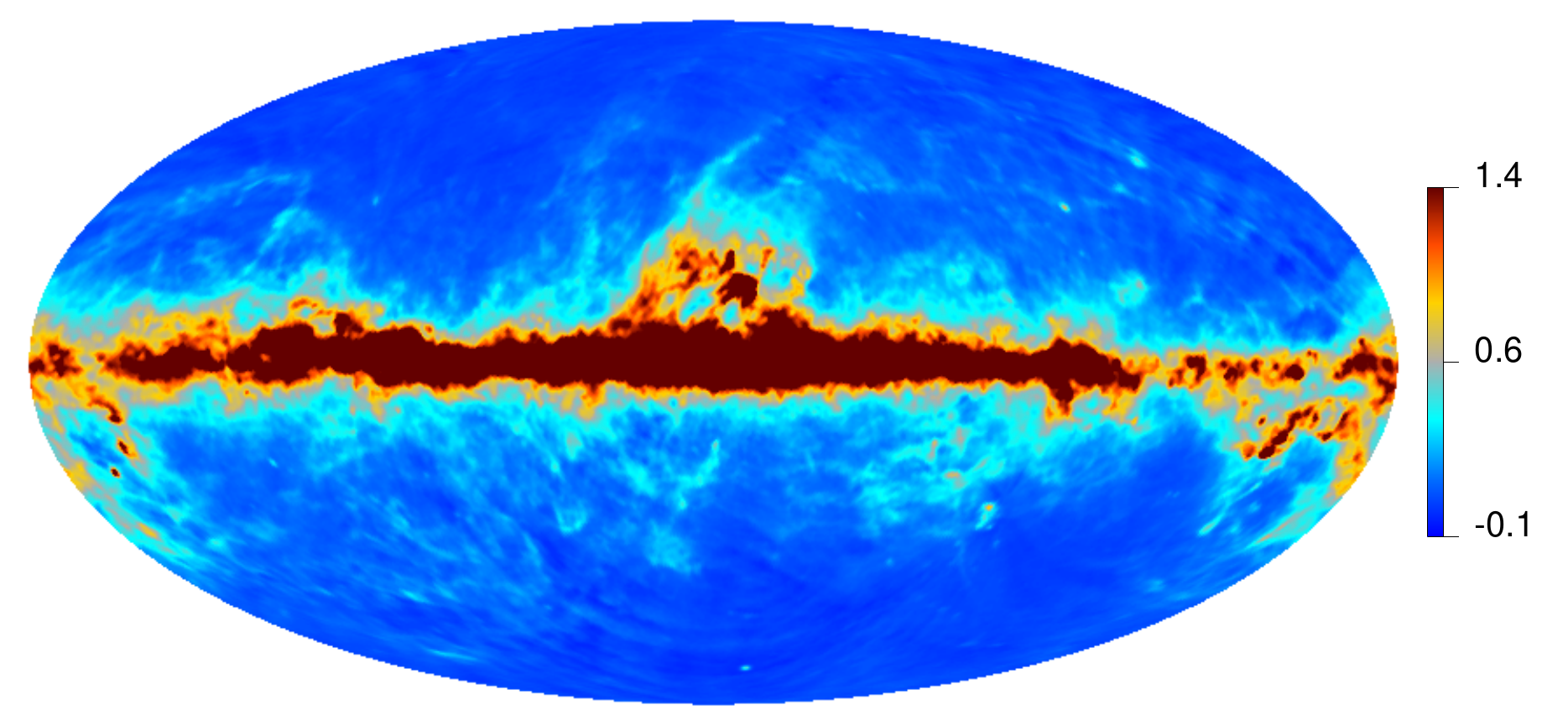}
\end{tabular}}
\caption{All-sky IRAS 12\,$\mu$m map in MJy/sr. Top: From the IRIS reprocessing \citep{MAMD2005}. Bottom: After removing the zodiacal residuals.}
\label{fig:allskyIRIS}
\end{figure}

The IRIS maps \citep{MAMD2005} were partially corrected for residual zodiacal emission compared to the original IRAS products \citep[ISSA plates --][]{Wheelock1993}. This was achieved by setting the emission at scales larger than $1^\circ$ to the one of DIRBE, which benefited from a better zodiacal subtraction \citep{Kelsall1998}. Nevertheless, the DIRBE and IRIS maps still show significant zodiacal residuals, especially at 12 and 25\,$\mu$m (see Fig.~\ref{fig:allskyIRIS}). Most of the zodiacal residuals are oriented parallel to the ecliptic plane, but because of the scanning strategy of IRAS, which followed the ecliptic plane, and because of the way zero levels were estimated in the ISSA plates products, there are also residuals perpendicular to the ecliptic plane. 

In order to improve on the zodiacal residual subtraction in the IRIS 12~$\mu$m band, we performed an adaptive filtering tailored to extracting residual emission oriented along or perpendicular to the ecliptic plane. The full method will be described in a forthcoming paper (Miville-Desch\^enes et al., in prep). Here we only summarise the processing steps. 

The main difficulty in estimating zodiacal residuals based on the map itself (and not returning to the original data timeline) is to separate it from the interstellar medium emission itself, which dominates the signal almost everywhere. To estimate the zodiacal map $Z(\vec{r})$ of an emission map $I(\vec{r})$ (here the IRIS 12~$\mu$m map), we performed a filtering $F$ of a residual map $R(\vec{r})$, $$Z(\vec{r}) = F(R(\vec{r})),$$ where $R$ is a difference map between the original map $I$ and a reference map $I_0$ (assumed to be free of zodiacal residuals) times a spatially varying correlation coefficient $\alpha$, $$R(\vec{r}) = I(\vec{r}) - \alpha(\vec{r}) I_0(\vec{r}).$$

Here the reference map we used is the 100~$\mu$m map, which combines the original IRIS map with the map of \cite{Schlegel1998}, which has fewer zodiacal residuals at large scales. This was also used in the all-sky dust model of \cite{PlanckCollaborationXI}. It is available in the Planck archive\footnote{\url{http://pla.esac.esa.int/pla/aio/product-action?MAP.MAP_ID=IRIS_combined_SFD_really_nohole_4_2048.fits}}.

The correlation coefficient $\alpha(\vec{r})$ was estimated benefiting from the healpix gridding of the data. Specifically, $\alpha$ was estimated on a coarser healpix grid (we used $n_{\rm side} = 32$), where for each large healpix pixel, the linear regression correlation coefficient was calculated using $n_{\rm side} = 2048$ pixels of $I$ versus $I_0$ encapsulated in the $n_{\rm side} = 32$ large pixels. This coarse map of $\alpha$ was then regridded to $n_{\rm side} = 2048$ to calculate $R(\vec{r})$. 
 
The filtering $F$ applied on $R$ was tailored to extract large-scale structures parallel and perpendicular to the ecliptic plane. To do this, we filtered $R$ using a median filtering technique with an elongated window. First a window parallel to the ecliptic plane was used with a width of $20^\circ$ in ecliptic longitude and only 3' in latitude (basically only two pixels). The map of $Z$ we derived in this way was removed from $I,$ and the whole process was reproduced using a perpendicular filtering window of 3' in longitude and $40^\circ$ in latitude. 

\begin{figure}[!t]
\centerline{\begin{tabular}{c}
\includegraphics[width=0.36\textwidth]{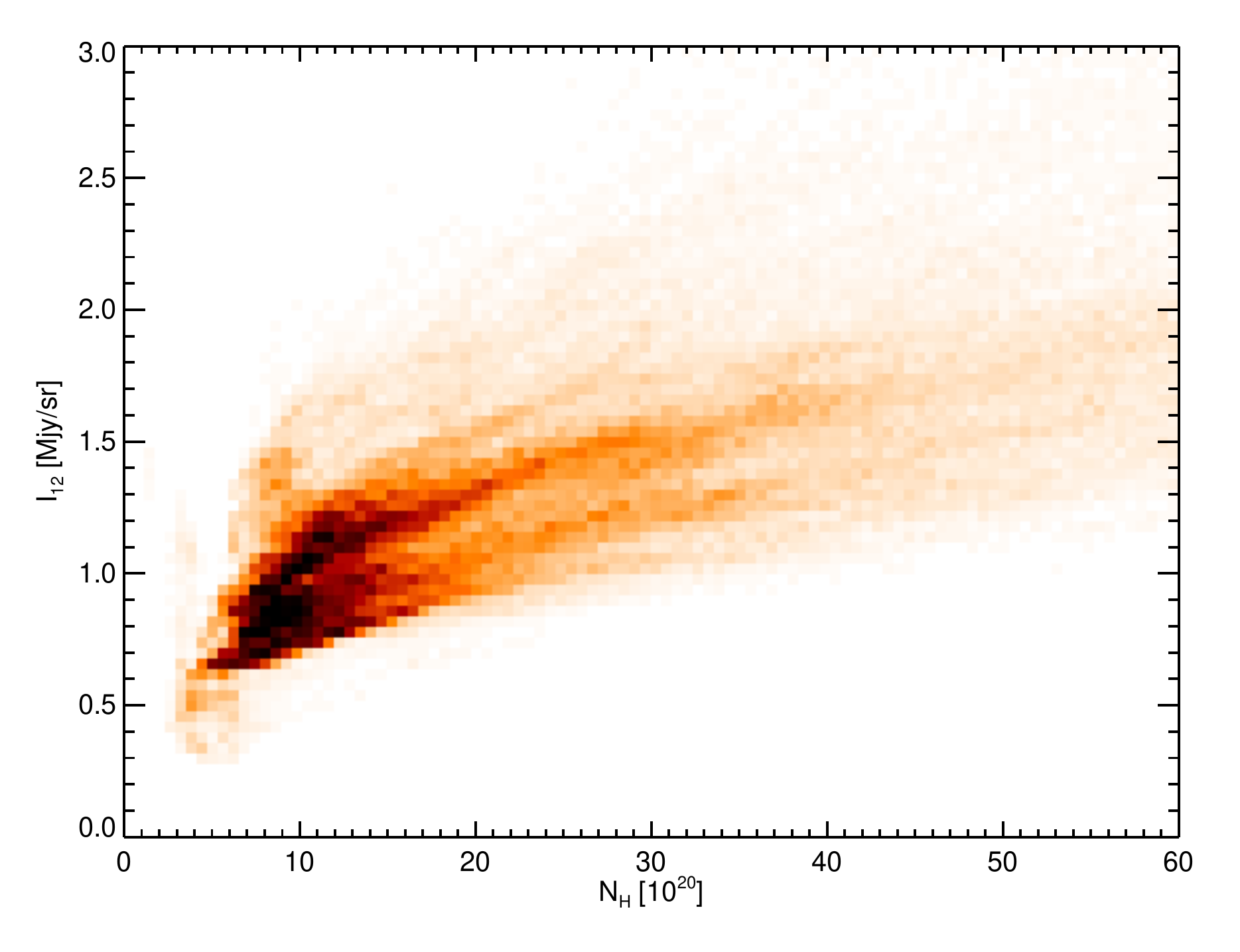} \\
\includegraphics[width=0.36\textwidth]{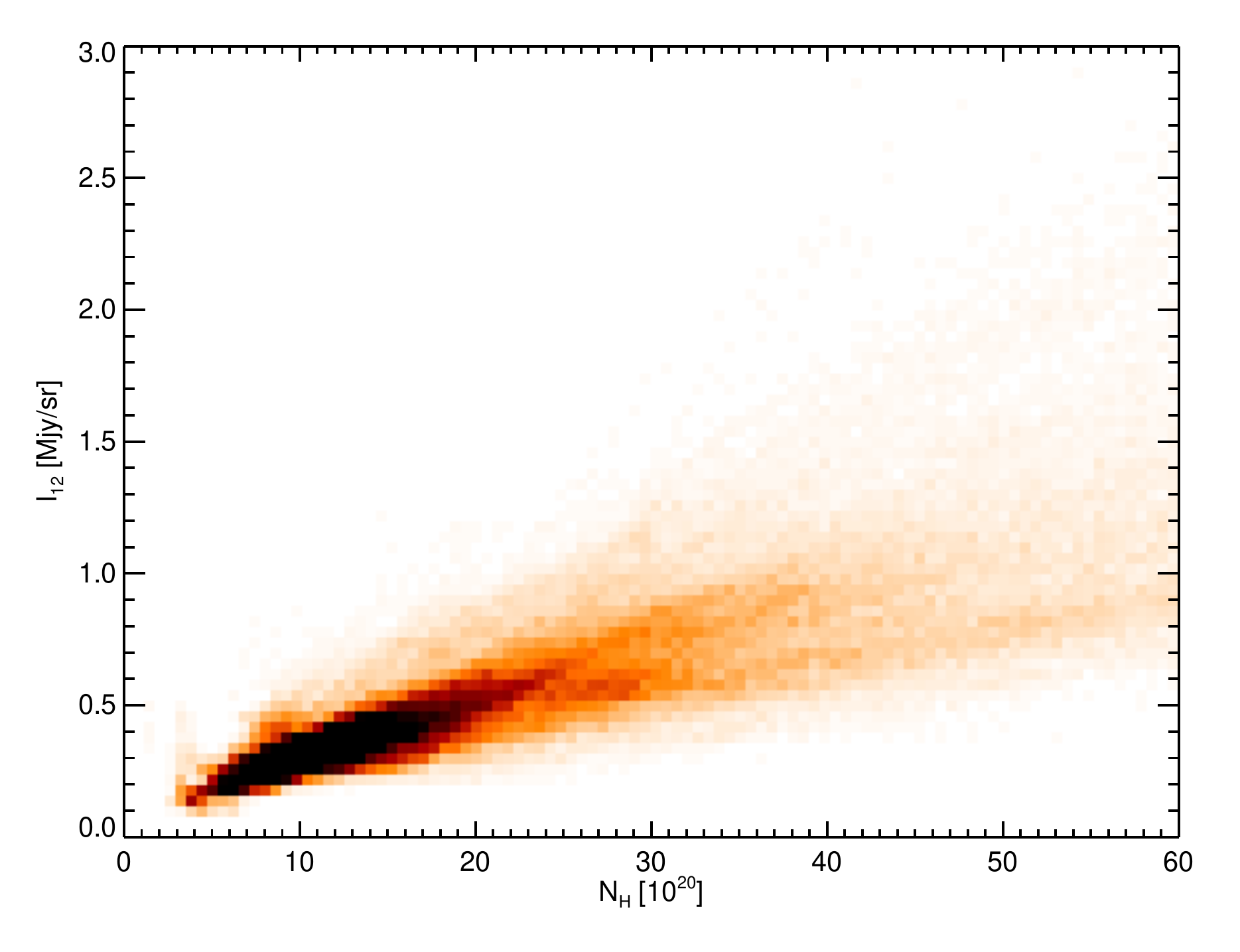}
\end{tabular}}
\caption{Two-dimensional histogram of the 12~$\mu$m emission as a function of $N_{\rm H}$ deduced from the $\tau_{353~GHz}$ dust optical depth map of \citet{PlanckCollaborationXI} before (top) and after (bottom) zodiacal correction. The data are from the maps used in the current paper; all maps are at $1^\circ$ resolution, and only the pixels in the mask shown in Fig.~\ref{figure_mask} were used to build the histograms.}
\label{fig:correl_12_tau}
\end{figure}

The resulting 12~$\mu$m map obtained after this process is shown in the right panel of Fig.~\ref{fig:allskyIRIS}. Faint zodiacal residuals are still visible, but the improvement is significant. It can be evaluated by considering the correlation of the 12~$\mu$m emission with other tracers of the interstellar medium column density. As an example, we present in Fig.~\ref{fig:correl_12_tau} the correlation of the original (top) and zodi-corrected (bottom) IRIS 12~$\mu$m maps with $\tau_{353~GHz}$ \citep[here converted into column density $N_{\rm H}$][]{PlanckCollaborationXI}, convolved at a resolution of $1^\circ$ and limited to the pixels we used in our analysis (see the mask shown in Fig.~\ref{figure_mask}). The correlation coefficient is increased from 0.82 to 0.86.

\end{appendix}

\end{document}